\documentclass[11pt,a4paper,headsepline,footsepline,BCOR12mm,DIV13]{article}

\usepackage[utf8]{inputenc}
\usepackage[margin=0.5in]{geometry}
\usepackage{todonotes}
\usepackage[pdftex,
            pdfauthor={Diblen et al.},
            pdftitle={Interactive web-based visualization of multi-dimensional physical and astronomical data},
            pdfproducer={Latex with hyperref},
            pdfcreator={pdflatex}]{hyperref}

\usepackage[T1]{fontenc}
\usepackage{authblk}

\usepackage{graphicx}
\usepackage{hyperref}
\hypersetup{
    colorlinks=true,
    linkcolor=blue,
    filecolor=magenta,      
    urlcolor=blue,
    citecolor=blue
}

\title{Interactive web-based visualization of multi-dimensional physical and astronomical data}
\author[2]{Faruk Diblen\thanks{f.diblen@esciencecenter.nl}}
\author[1,3]{Luc Hendriks\thanks{luc.hendriks@gmail.com}}
\author[1]{Bob Stienen\thanks{b.stienen@science.ru.nl}}
\author[1,3]{Sascha Caron\thanks{scaron@nikhef.nl}}
\author[2]{Rena Bakhshi\thanks{r.bakhshi@esciencecenter.nl}}
\author[2]{Jisk Attema\thanks{j.attema@esciencecenter.nl}}
\affil[1]{High Energy Physics, IMAPP, Radboud University Nijmegen, Heyendaalseweg 135, 6525 AJ Nijmegen, NL}
\affil[2]{Netherlands eScience Center, Science Park 140, 1098 XG Amsterdam, The Netherlands}
\affil[3]{Nikhef, Science Park 105, 1098 XG Amsterdam, N}

\begin{document}
\maketitle

\begin{abstract}

\section*{}
In this manuscript, we propose to expand the use of scientific repositories such as Zenodo and HEP Data, in particular in order to better examine multi-parametric solutions of physical models.
The implementation of interactive web-based visualizations enables fast and comfortable re-analysis and comparisons of phenomenological data.

In order to illustrate our point of view, we present some examples and demos for dark matter models, supersymmetry exclusions and LHC simulations.

\vspace{1em}

\noindent
\textbf{Keywords:} visualization, multi-dimensional data, machine learning, simulations, particle physics
\end{abstract}


\section{Introduction}\label{sec:intro}

Practically any research done in modern physics nowadays is grafted on (simulated) data. Whether it is a the investigation of the Higgs boson properties or the search \& exclusion of new models for physics beyond the Standard Model at the LHC, the investigation of gravitational waves or the identification of dark matter. In all these scientific efforts the exploration of data with the help of physical models plays a key role.
The models are often complex, i.e. they depend on various physical parameters and their interpretation may depend on systematic effects described with the help of additional nuisance parameters.

Traditionally scientific data\footnote{Before the area of histograms there was the time of pictures, e.g. from bubble chambers.} is provided by the experiments mainly in the form of 1-dimensional histograms and data analysis typically required a comparison of the model to the histogram of data. The scientific models investigated were also of low complexity.
Predictions of models that describe physics are typically compared to the data using often time-consuming simulations of the underlying physical processes for a large number of model parameter sets.
Finally, the best-fit contours of the models are  presented in the form of  likelihoods or posterior distributions as functions of a model parameter $\theta$, typically in the form of 1-2 dimensional figures in scientific publications. 

In the field of searches beyond the Standard Model typically 95\% confidence level upper limits are provided, i.e. model parameter sets are classified between "excluded" and "allowed".

\begin{figure}[!htbp]
    \centering
    \includegraphics[width=53mm]{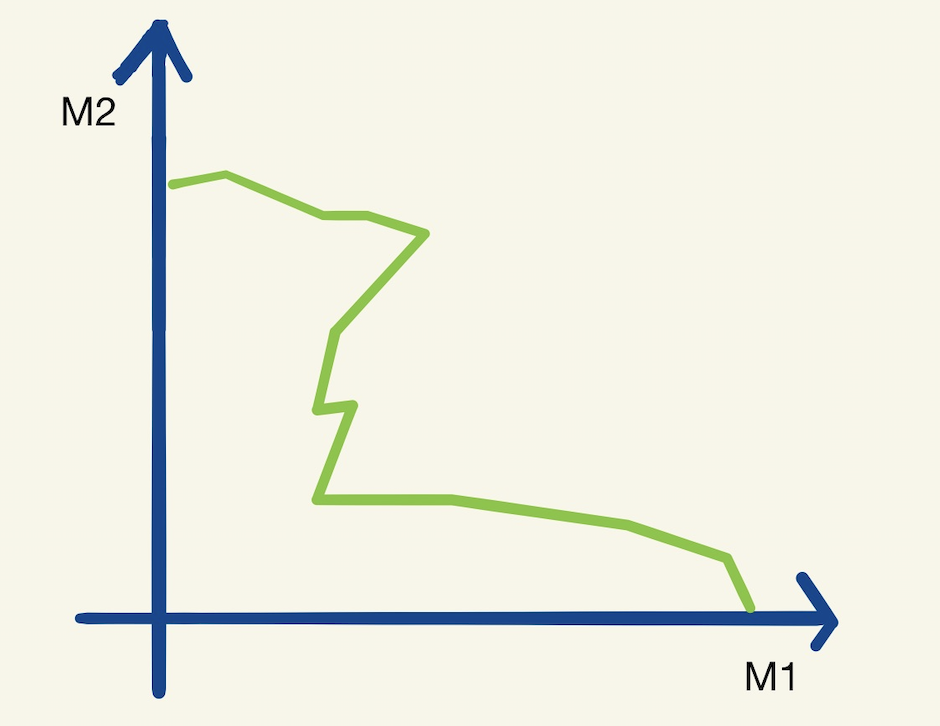}
    \includegraphics[width=50mm]{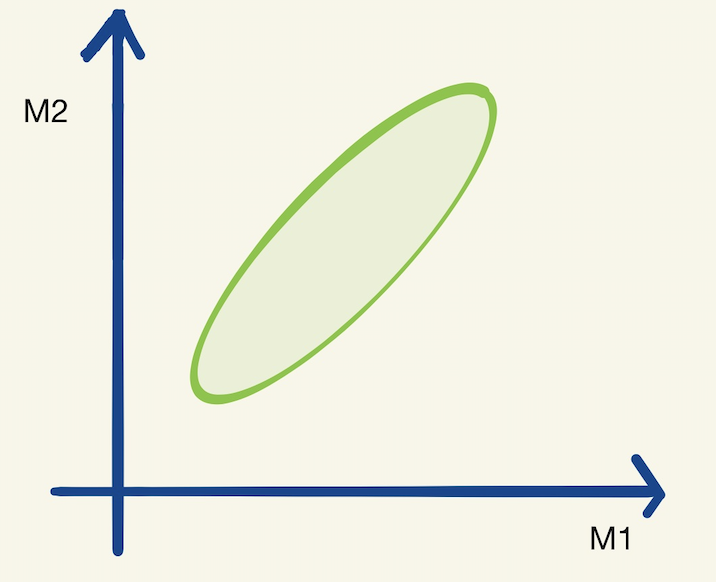}
    \caption{A typical exclusion limit curve (left) and a typical maximum likelihood interval (right) as a function of the two parameters M1 and M2. Let us assume that the hypothetical real physical model has 20 parameters M1-M20.}\label{fig:exclusion}
\end{figure}

In order to visualise high-dimensional data, one can use 1- or 2-dimensional projections (e.g. some parameter variables have been marginalized or set to best fit values), or slices of the full space, but this comes at the cost of information contained in the visualisation itself.
An example of such a 1-dimensional exclusion contour and a maximum likelihood interval is given in Figure~\ref{fig:exclusion}.

The problem becomes even more apparent in the combination and comparison of the results of two different research data sets.  
To actually visualise data and make a comparison between two data sets, expert knowledge on the creation of those sets is often needed.

There are exciting visualization tools that solve these problems, but these are not open source~\cite{powerbi,ahlberg1996spotfire,plotly,heer2008graphical}, require technical expertise or are generic tools~\cite{superset,metabase}.

An open and user-friendly visualization tool would also allow researchers with limited-time (and limited technical skills) to find new correlations in published data. We believe that such a web-based and easy-to-use data-visualisation tool can generate new ideas and accelerate science.

Repositories like Zenodo~\cite{zenodo} allow to  store  likelihood or exclusion boundaries for thousands to millions  different parameter sets. 
Zenodo is today also widely used to publicly store the results of simulations used in scientific publications.

This manuscript proposes to build two additions to this new way of multivariate data explorations and shows working demonstrations.
\begin{itemize}
    \item \textbf{Interactive visualisation of data samples:} We believe that there should also be an easy way to visualize, explore and compare multi-dimensional data. This allows a quick and intuitive  visualisation and analysis of the model data. We propose a tool for ``online'' $1-3$ dimensional  visualisation and histogramming of high-dimensional data sets which could be connected to online repositories such as Zenodo.
    \item\textbf{Generalization of data samples:} As described in~\cite{Brooijmans:2020yij}, the regression and classification with Machine Learning (ML) allows a practical interpolations \textit{in between} the provided model solutions. ML-based interpolations are best practice also for high-dimensional models. Such tools are built by the community (or could be automatically build on some of the Zenodo data sets). This would allow to determine exclusion, likelihood, posteriors etc. for an arbitrary set of model parameters.
\end{itemize}

In the following we discuss a few example cases and provide links to working demos build on 
on SPOT~\cite{Diblen:2018mib, attema_jisk_2020_4116521} with data sets available on Zenodo via phenoMLdata.org.

\section{A demonstration of interactive visualization}\label{sec:the-why}

\texttt{phenoMLdata} is an online interactive plotting interface to a database of publicly available data sets, based on the open-source SPOT framework. A prototype of the website can be found at \href{http://spot.phenomldata.org}{http://spot.phenomldata.org}. 

The tool allows for the creation of histograms, line plots, pie charts and scatter plots (both 2-dimensional and 3-dimensional) of any (combination) of the variables without downloading the data to user's computer. As conventional with plotting tools the properties of these plots, like ranges of the axes, colours and points sizes can be customized and made data dependent.

During the creation of a plot, all available data from the data sets is plotted. SPOT adds interactivity between the plots by linking them. In a session with a histogram and a scatter plot users can for example select a region in the scatter plot. The histogram then automatically alters itself to only show data contained in the user's selection, making it possible to make real-time filtering on the data.

Another advantage is the possibility to compare different data sets. The online interface provides access to a database containing the data sets. Any data set in this database can be selected for visualisation. By selecting multiple data sets it is possible to plot (the same) variables of different data sets in the same plot. This makes it possible to compare e.g. exclusion boundaries of different papers, projected onto any available plane.

Any visualisation made with the tool can be exported as a provenance file (active session) for sharing. Any other user that uploads this file can then use the visualisations for themselves and has access to the full interactive arsenal of SPOT in doing so. The next section showcases this explicitly: any of the provided examples comes with a URL to the provenance file for that specific example. Uploading this file to SPOT opens the full interactive version of said example.

\section{Examples}\label{sec:examples}
For each example we provide the "session URL" in the Figure captions. All the session files can be found on \href{https://doi.org/10.5281/zenodo.4247860}{https://doi.org/10.5281/zenodo.4247860}.

\subsection{Example 1: SUSY-AI}\label{sec:ex1}

A primary goal of particle physics is to find signals that are predicted by extensions of the Standard Model. These extensions always come with several new parameters, such as e.g. the masses and couplings of new particles.
The ATLAS experiment provides in~\cite{Aad2015} for a 19-dimensional model of new physics (the so called pMSSM) and about 310,000 realizations of this model (with randomly selected parameter sets) the information whether this parameter set is excluded (or not) by ATLAS measurements.
 These  310,000 model configurations with binary exclusions  allowed the construction of a Machine Learning classifier to predict the exclusion contour in the full 19-dimensional model space~\cite{Caron2017}.
The investigation of the exclusion of the 19-dimensional model parameters is, nevertheless, typically still done through two-dimensional projections.

\begin{figure}[!htbp]
    \begin{center}
        \includegraphics[width=0.7\textwidth]{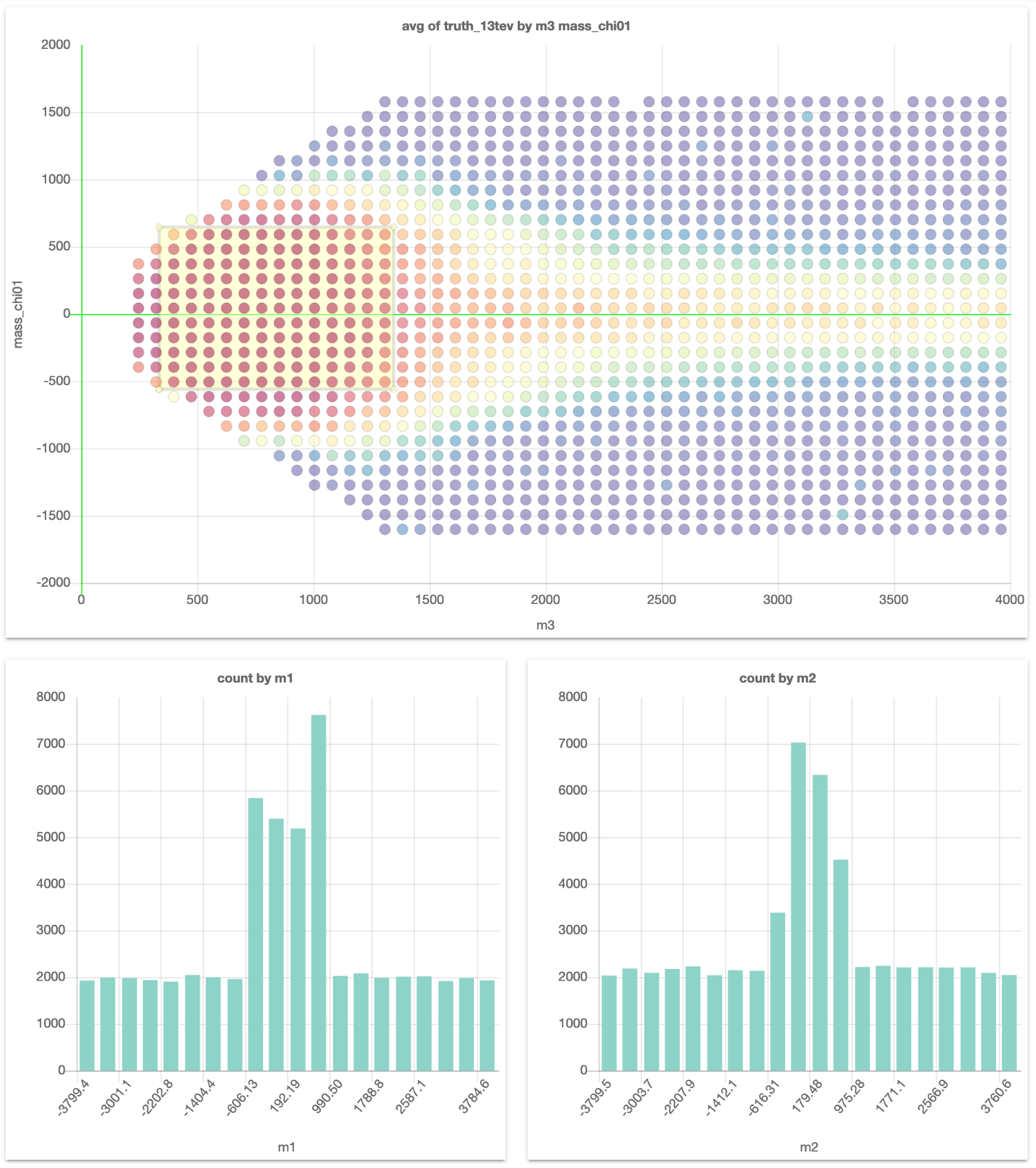}
    \end{center}
    \caption{The example projection made of a subset of the data in~\cite{Aad2015}. The graph on top shows the projection on 2 of the 19 free parameters and the colors indicate the average exclusion for the model configurations in each for the bins. The two graphs on the bottom can be used to make cuts on two of the remaining 17 parameters. Interactive dashboard of these graphs can be accessed \href{http://spot.phenomldata.org/\#session=https://s3-eu-west-1.amazonaws.com/www.phenomldata.org/session_susyai.json}{here}.}\label{fig:susyai}
\end{figure}

However, storing the data in a database with an interactive plot tool would solve this problem. The reader can then simply take the actions that interest them.
In this example, a subset of the data from~\cite{Aad2015} is saved so that every projection and every slice can now be plotted.
Figure~\ref{fig:susyai} shows (on the right) an example of an exclusion plot that can be made, in which colours indicate the average exclusion in each bin. Using the dynamic links between the plots in this figure, one can make on-the-fly slices in the project plot using the plots on the right, allowing for a quick and full exploration of the high-dimensional data set. This shows that a tool like phenomldata.org could be used to accompany papers with an effectively unlimited number of (exclusion) plots and projections based on the data used in writing the paper.

\subsection{Example 2: Galactic Center excess model solutions}
\label{sec:ex2}
In~\cite{Achterberg:2017emt} and \cite{Caron:2015wda} a 19-dimensional theoretical model \footnote{A version of the pMSSM model also used in Example 1 (Section 3.1).} was tested against a measurement of an excess in the Galactic Center (GC), which was observed during the analysis of gamma rays.

In this work, a very small area of the parameter space was found that could explain this excess and lead to a particular candidate for dark matter.
In these papers, various scatter plots are shown to highlight where in the parameter space these solutions are. However, these various two-dimensional plots are merely slices in the 19-dimensional parameter space. From a theoretical standpoint we know that this parameter space is very complex: it contains delta peaks, step functions and high dimensional correlations. The 2D diagrams therefore do not show the complete information contained in 19D space.
In addition, mapping all the combinations of 19D space would require 171 scatter plots, and this still does not show the correlations shown in more than two dimensions.

Another researcher might be interested in a scatter plot or histogram that is not in the paper, because this person needs it to design an experiment to verify if the dark matter candidate exists. As this plot is not in the paper, this person needs to contact the authors and hope that they still have the results of this paper in high-dimensional format.

\begin{figure}[!htbp]
    \begin{center}
        \includegraphics[trim=0 0 0 0,clip,width=0.7\textwidth]{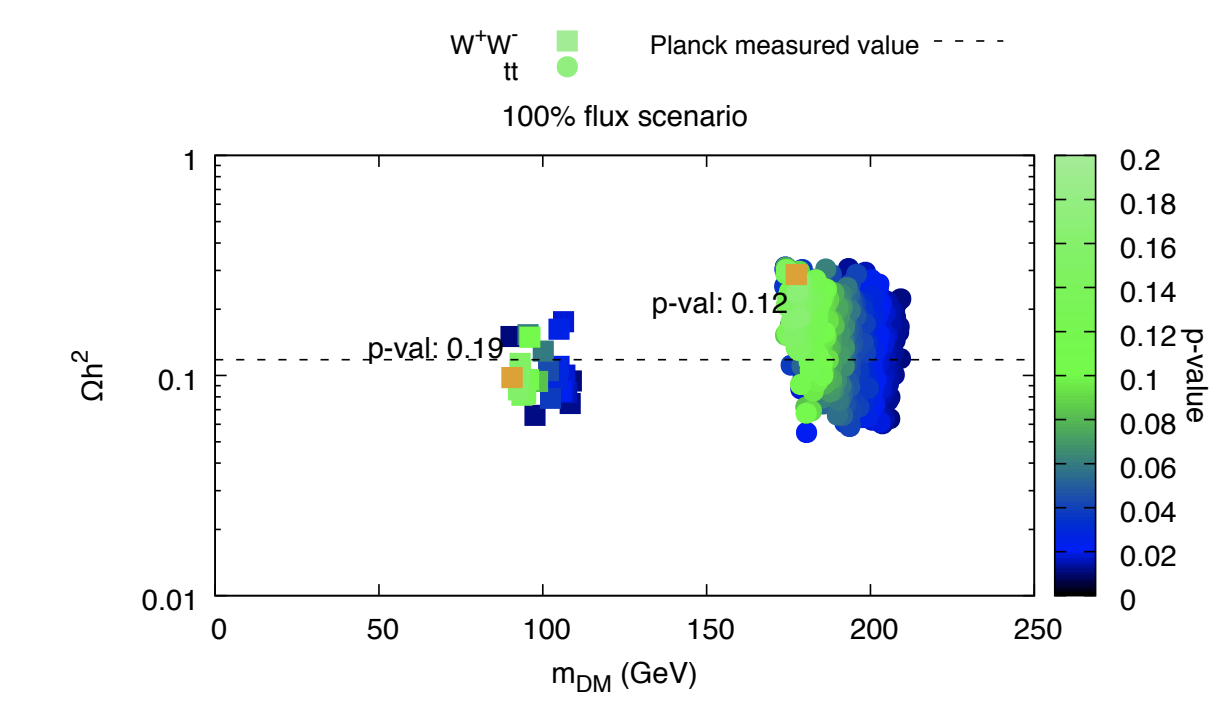}
    \end{center}
    \caption{Copy of Figure 6a in~\cite{Achterberg:2017emt}. This plot has been recreated in SPOT and a screenshot is shown in Figure \ref{fig:screenshot_gc_excess}. The figure is the same as the plot in the top left.}\label{fig:screenshot_gc_comparison}
\end{figure}

\begin{figure}[ht]
    \begin{center}
        \includegraphics[width=0.9\textwidth]{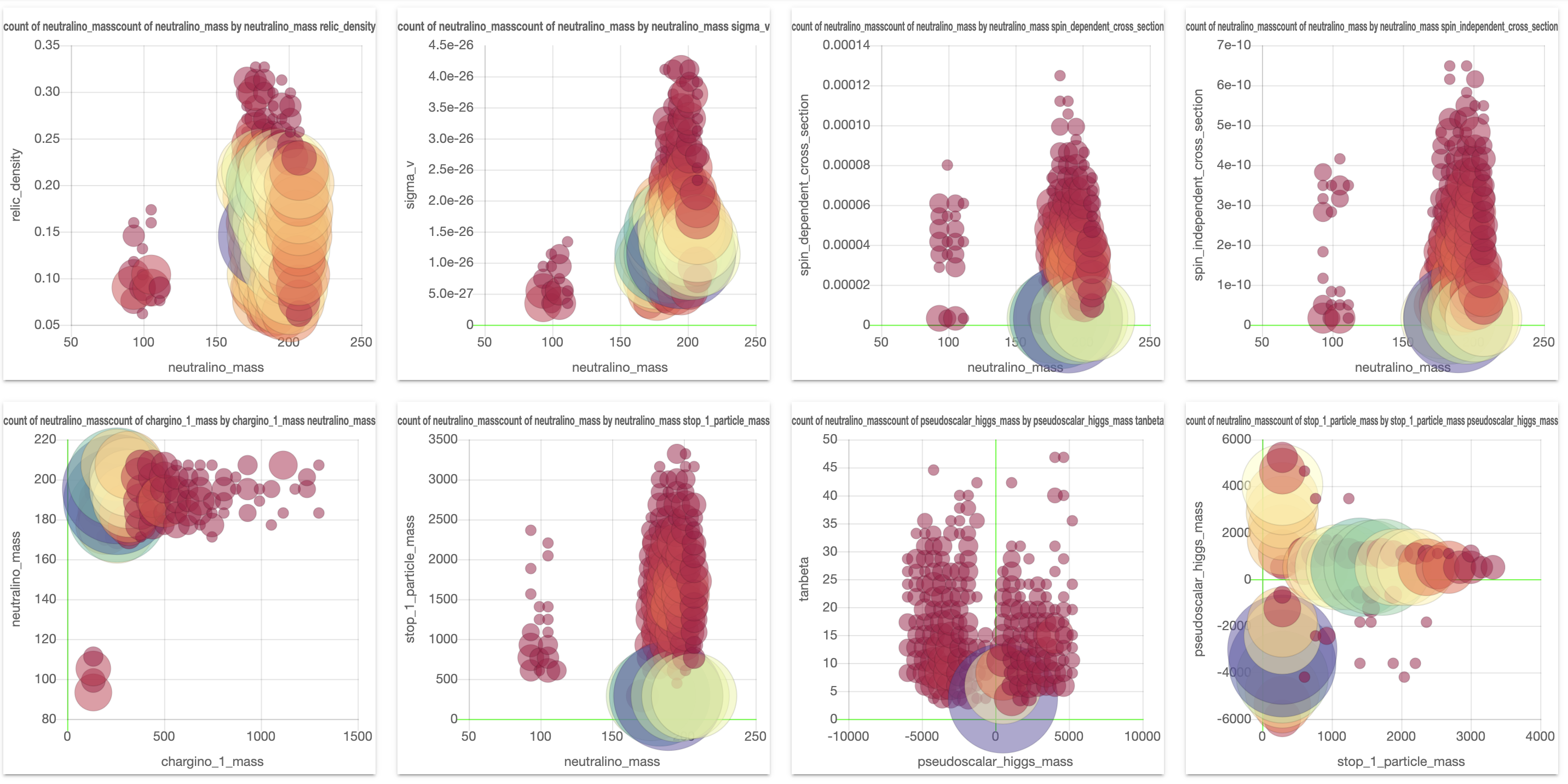}
    \end{center}
    \caption{Recreation of the plots in~\cite{Achterberg:2017emt} using of SPOT. The interactive dashboard of this data set can be accessed \href{http://spot.phenomldata.org/\#session=https://s3-eu-west-1.amazonaws.com/www.phenomldata.org/session_1709.json}{here}.}\label{fig:screenshot_gc_excess}
\end{figure}

Additionally, while different theoretical models will have different input parameters, the output parameters will most likely overlap. For example, all dark matter models must predict the mass and cross section of the dark matter candidate, and all models that match the GC excess derive a likelihood.
By focusing on these overlapping parameters, one can compare results of different theoretical models in the same plots and have a very fast and convenient way to compare results of different papers. An example of the recreation of the plots in~\cite{Achterberg:2017emt} can be found in Figure~\ref{fig:screenshot_gc_excess}. An example of two plots in~\cite{Achterberg:2017emt} can be found in Figure~\ref{fig:screenshot_gc_comparison}.

\subsection{Example 3: LHC collider simulation events and humans finding signals}
\label{sec:ex3}
A benchmark data set containing $>10^8$ simulated high-energy collision data has been provided by participants of the www.darkmachines.org initiative and the 2019 Les Houches workshop~\cite{Brooijmans:2020yij}.
The generated LHC events correspond 
to a center-of-mass energy of 13~TeV. 
Events for the background and signal processes are generated using the event generator MG5$\_$aMC@NLO~v6.3.2~(Madgraph) and versions above~\cite{Alwall:2014hca}. Also a quick detector simulation was performed and the high-level objects like jets, b-jets, electrons, muons and photons have been reconstructed. The charge, object type and 4-vector (energy $E$,  transverse momentum $p_T$, pseudorapidity $\eta$ and azimuthal angle $\phi$) are stored for each object. A description of the requirements and the data structure can be found in~\cite{Brooijmans:2020yij}.

This data set can be useful for various phenomenological studies. One of the goals is to develop and compare new strategies for searching for signals from new physics. 
Here it is interesting to see which regions of phase space are selected by new signal detection algorithms. Cuts of such Machine Learning based algorithms are typically represented by a complicated multidimensional hyper- surface of the 4-vectors and objects.
An online interactive multidimensional visualisation tool such as SPOT would allow to quickly compare e.g. where such cuts should have been made (or where such cuts where made if the algorithms output can be included/uploaded into the data set to be visualised).  

\begin{figure}[!htbp]
    \begin{center}
        \includegraphics[width=0.85\textwidth]{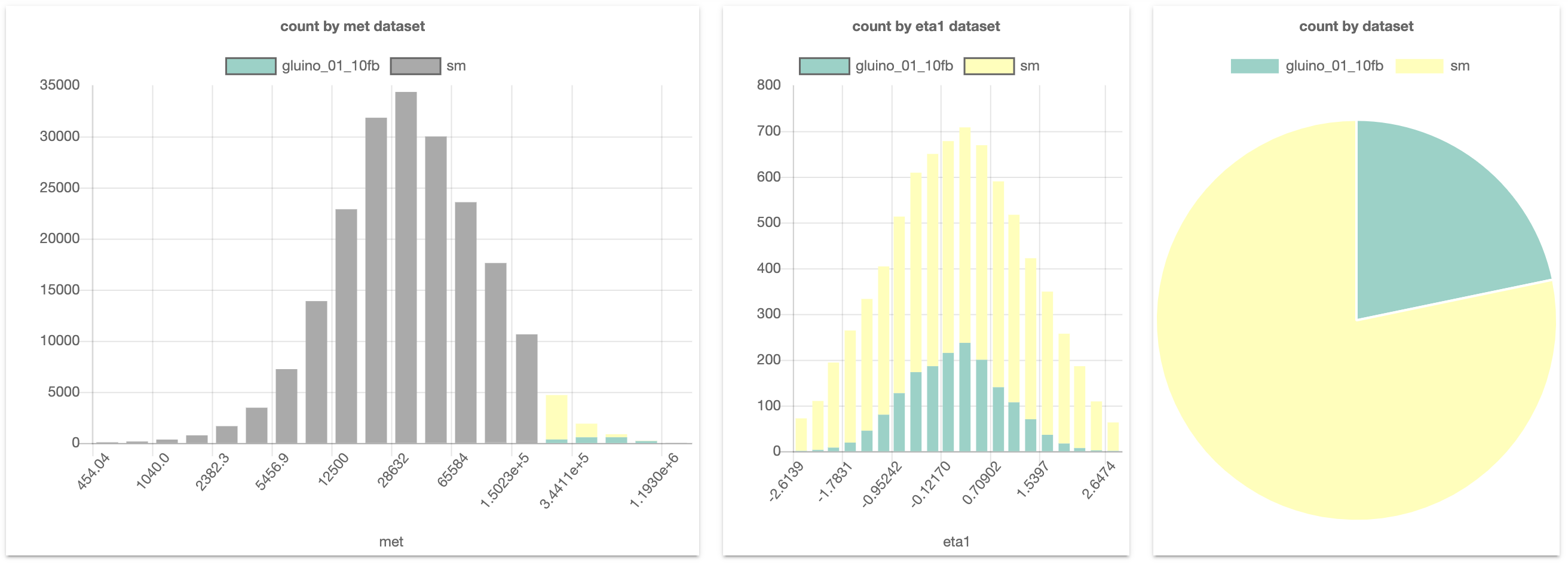}
    \end{center}
    \caption{SPOT-based comparison of a possible LHC signal (here a gluino production from Supersymmetry) and background. The interactive dashboard of this data set can be accessed  \href{http://spot.phenomldata.org/\#session=https://s3-eu-west-1.amazonaws.com/www.phenomldata.org/session.json}{here}.}\label{fig:screenshot_LHC}
\end{figure}

An example of a SPOT session with this data can be found in Figure~\ref{fig:screenshot_LHC}. Here a comparison is made between a possible gluino signal from Supersymmetry and the expected background events. All events are from simulation.

It would be interesting to create a challenge where people can search for the signal by applying selection criteria by hand. Would they be better than machine learning-based anomaly detection algorithms? A signal region found by a "human" could be used as a "data derived" signal region in an independent data set, similar to that proposed by ATLAS in~\cite{Aaboud:2018ufy}.

\subsection{Example 4: Fermi Point Source catalogue}
\label{sec:ex4}
The Fermi FL8Y Point Source catalogue~\cite{FL8Y_Dataset} contains a list of found point sources using eight years of Fermi data. It is a big table containing locations and properties of various types of point sources. If one would be interested to quickly check where the most unresolved point sources are, the catalogue has to be downloaded and then you have to write a visualisation script to filter out the unresolved point sources and plot the latitude and longitude of the corresponding rows. This requires technical knowledge and is quite time-consuming for such a simple check.

\begin{figure}[!htbp]
    \begin{center}
        \includegraphics[width=0.8\textwidth]{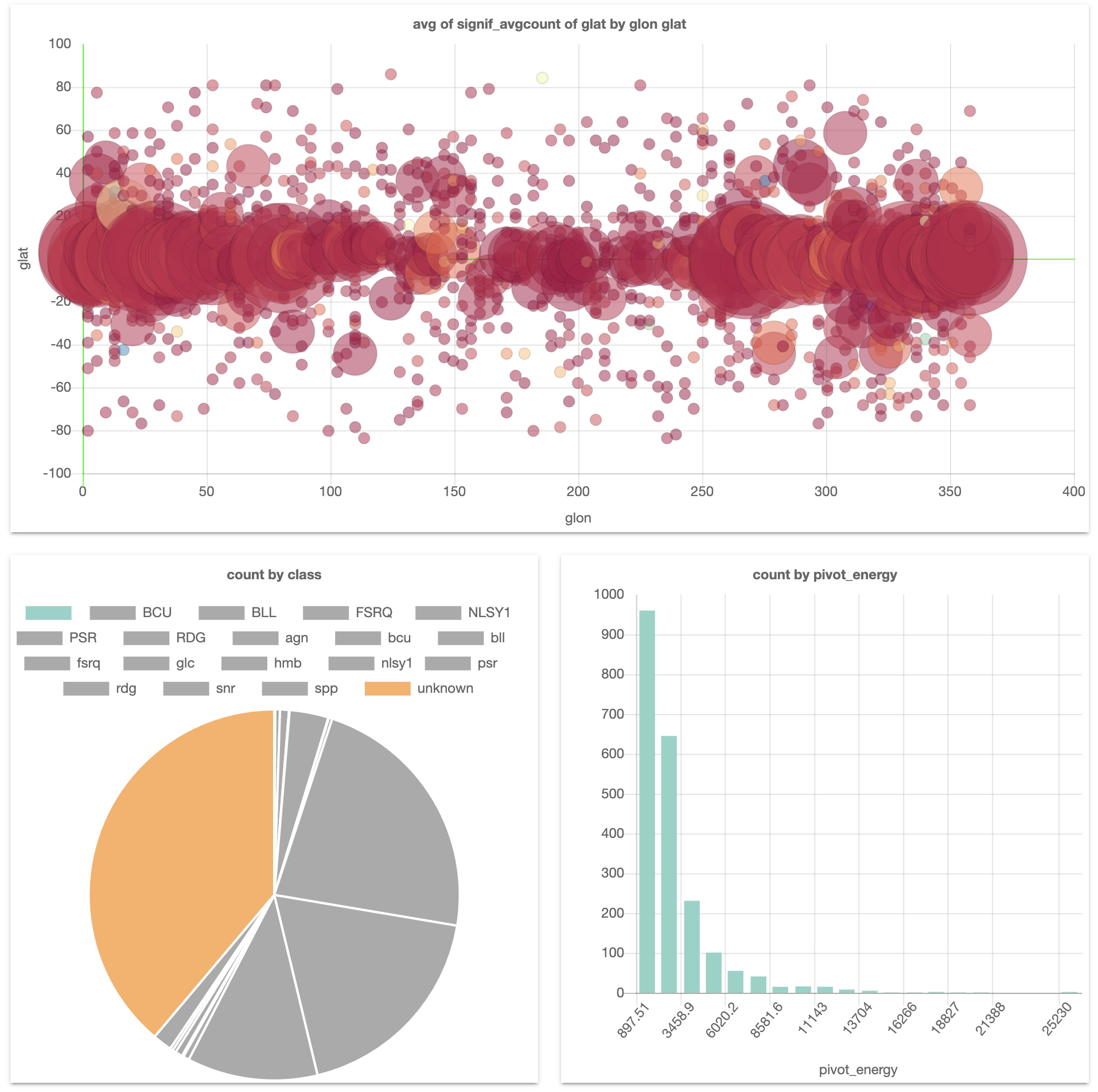}
    \end{center}
    \caption{SPOT visualisation of the FL8Y Fermi Point Source catalogue~\cite{FL8Y_Dataset}. The interactive dashboard of this data set can be accessed \href{http://spot.phenomldata.org/\#session=https://s3-eu-west-1.amazonaws.com/www.phenomldata.org/session_fl8y.json}{here}.}\label{fig:screenshot_fermi}
\end{figure}

Alternatively, in SPOT it requires only a few clicks to generate these plots and conclude they lie mainly in the Galactic Plane (where also most of the diffuse background radiation is). An example visualization of this data set can be found at \ref{fig:screenshot_fermi}.

\section{Conclusion}\label{sec:conclusion}

In this note, we propose the expansion of scientific repositories such as Zenodo to allow easy web-based visualization of data. We show some examples where such visualization could speed up science.
Additionally, it would be an important step to encourage the HEP community to study physical models and publish results in their full dimensionality.
This would allow and encourage a revision of the results with different model parameters, the search for anomalies (or errors) in the published data, the generalization of the results with machine learning and a better comparison of the scientific publication.

We would like to emphasize that the development and the maintenance of such a tool must be a collaborative effort. We hope the community will realize the importance of our solution so that we can build these tools together.

\section*{Acknowledgments}
This work is supported by the Netherlands eScience Center under the project \href{https://www.esciencecenter.nl/projects/idark}{iDark: The intelligent Dark Matter Survey}.

\bibliographystyle{abbrv}
\bibliography{frontiers_references}

\end{document}